%% file: article_KID.tex
\pdfoutput=1
\documentclass{JINST}
\usepackage{textcomp} 

\DeclareGraphicsExtensions{.pdf,.png,.jpg} 
\title{Electronics and data acquisition demonstrator for a kinetic inductance camera}
\author{O.~Bourrion$^a$\thanks{Corresponding author.}, A.~Bideaud$^b$, A.~Benoit$^b$, A.~Cruciani$^b$, J.F.~Macias-Perez$^a$, A.~Monfardini$^b$,
M.~Roesch$^c$, L.~Swenson$^b$, C.~Vescovi$^a$.\\
\llap{$^a$}Laboratoire de Physique Subatomique et de Cosmologie,\\ 
Universit\'e Joseph Fourier Grenoble 1,\\
  CNRS/IN2P3, Institut Polytechnique de Grenoble,\\
  53, rue des Martyrs, Grenoble, France\\
\llap{$^b$}Institut Néel, CNRS/UJF, \\
  25 rue des Martyrs, Grenoble, France \\
\llap{$^c$}Institut de RadioAstronomie Millimétrique \\
  300 rue de la Piscine, 38406 Saint Martin d'H\`eres, France \\
}  

\abstract{
A prototype of digital frequency multiplexing electronics allowing the real time monitoring of microwave kinetic inductance detector (MKIDs) arrays for mm-wave astronomy has been developed. It requires only 2 coaxial cables for instrumenting a large array. For that, an excitation comb of frequencies is generated and fed through the detector.
The direct frequency synthesis and the data acquisition relies heavily on a large FPGA using parallelized and pipelined processing. The prototype can instrument 128 resonators (pixels) over a bandwidth of 125\,MHz. This paper describes the technical solution chosen, the algorithm used and the results obtained.
} 


\begin{document}


\input{intro}
\input{det_description}
\input{instrumentation}

\input{conclusion}


\include{biblio}
\end{document}

%% file: intro.tex
\section{Introduction}

Millimeter and submillimeter observations have proven  to be of primary importance for a wide range of astrophysical fields
going from the study of star forming regions in the Galaxy (\cite{ward2007}) to the measurement of the cosmic microwave background
(CMB) temperature and polarization anisotropies (see for example \cite{archeops,planck2011}).  
This has motivated a fast development of new technologies to improve the sensitivity of individual detectors
to match the needs of new generation of instruments.  For example, there is great interest in the community for the measurement of
the primordial CMB B modes in order to detect primordial gravitational waves.  These modes are expected to be extremely
faint in most cosmological scenarios (see for example \cite{efstathiou}) and in consequence high precision measurements are needed. 

The currently used detectors, mainly bolometer-based, are
limited by photon noise and therefore no further improvement in their sensitivity can be achieved. The only way to improve
the overall instrumental sensitivity is to increase the number of detectors in the focal plane.  In the past decade, arrays of
hundreds of individual bolometers have dominated continuum submillimeter and millimeter astronomy (eg. MAMBO2, BOLOCAM).
Full-sampling arrays with up to thousands of pixels are now reaching maturity, offering increased mapping speed and decreased
per-pixel manufacturing costs (eg. Apex-SZ, SPT, SCUBA2, LABOCA). Despite these great advances in technology, further
array scaling is strongly limited by the multiplexing factor of the readout electronics.

A very promising alternative to traditional bolometers are the microwave kinetic inductance detectors (MKID). MKIDs consist of a high-quality superconducting resonant circuit
electromagnetically coupled to a transmission line which are designed to resonate in the microwave domain \cite{Day,Mazin2,DoyleThesis}. Typically, for astronomical applications, resonances lie between 1 to 10\,GHz and have loaded quality factors exceeding $\rm Q_{L}=10^5$, corresponding to a typical bandwidth of $\rm \Delta f = f/Q_{L} \sim  10-100$\,kHz.  As the MKIDs resonant frequency can easily be controlled during manufacturing, it is possible to
couple a large number of MKIDs to a single transmission line without interference. This translates naturally in a single frequency-based multiplexing system
per measuring cable. Current frequency-multiplexing electronics are limited to hundreds of pixels because of their
reduced bandwidth. However, they should be easily improved to deal with thousands of pixels as expected in future experiments.


%% file: det_description.tex
\section{Detector description and instrumentation methodology}
\label{detector}
 In this paper a frequency-multiplexing electronics designed for the NIKA camera (\cite{Monfardini, Monfardini2011}) that is based on arrays
 of MKID detectors is described.
It is a dual-band millimeter-wave MKID camera for the IRAM 30-meter telescope aiming to assess the viability of MKIDs
for terrestrial astronomy. NIKA consists of a custom-designed dilution cryogenic cooler with a base temperature of $\sim$70\,mK
\cite{dilution}, an optical system and two MKID based arrays at 150 and 220\,GHz.
These arrays are respectively composed of 144 LEKIDs (Lumped Elements KIDs) and 256 antenna-coupled MKIDs.
Each MKID consists of a high quality factor superconducting resonator presenting a well-defined resonance
peak at several GHz (from 1 to 10\,GHz) with a few kHz bandwidth. The center frequency of the MKIDs resonance peak has been proved to be linearly proportional
to the sky incident power \cite{SwensonPhonon}. In other words, changes on the incident sky power produce a frequency shift of
the resonance peak which can be measured by monitoring the MKID response at a given single frequency. Indeed, the frequency
shift of the resonance peak  will lead to variations in the amplitude and phase of the MKID response at the monitored frequency \cite{MazinThesis}. 

Currently, hundreds of MKIDs can be coupled to a single
transmission line by spacing their resonance peaks by 1 to 2\,MHz \cite{Monfardini} in order to minimize interferences and crosstalk.
Thus, the incident sky power can be simultaneously measured in all detectors by monitoring each MKID at its resonant frequency.
Therefore, in order to readout the MKID array, a frequency comb having all its tones adjusted to the different resonators center frequencies must be injected in the detector. At the output of the transmission line coupled with the MKIDs, the resultant signal must be acquired and analyzed in order to determine the time dependent phase and amplitude variation of each sinusoids.

Two different solutions can be implemented in order to perform these operations.
The first obvious one is based on the Fast Fourier Transform analysis (FFT). In that case, the comb of frequencies, composed of summed sinusoids, is calculated off-line, stored in a circular buffer and used as the MKID array excitation signal. Then, the signal having interacted with the MKIDs is acquired and analyzed either off board by dedicated computers \cite{Yates} or on board by means of DSPs or FPGA \cite{ROACH}. In the first implementation a small size FPGA coupled to memories (stimulation and acquisition buffers) is sufficient and allows to easily cover a large bandwidth as the hardware processing power is reduced to a minimum. In the second case, a larger FPGA coupled to buffers holding stimulus and featuring a FPGA adapted FFT algorithm (channelizer) is used in order to maintain an acceptable readout rate (up to a 100\,Hz with 131\,k frequency bin).

Depending on the application targeted, the limitation of the FFT method may come from the number of points used. 
The SNR, which is a function of the amount of white noise spectral power integrated with the useful signal power, depends on the bandwidth and thus on the FFT size. 
 Another limitation of this method could be the output data rate, i.e. the time taken by an acquisition and analysis cycle.
The latter is particularly harmful in aerospace application where the MKID array can suffer glitches due to direct detection of cosmic ray. In such an eventually, the averaging introduced by the FFT could smear the glitch into the data and make it look like genuine data of interest. A faster readout is therefore an asset and mandatory in this application.

\begin{figure}[th]
\begin{center}
\includegraphics[angle=-90,width=7cm]{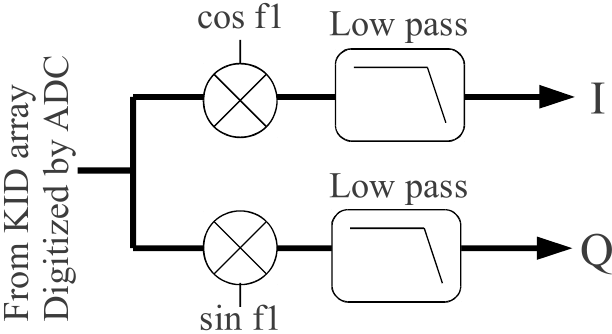}
\caption{Overview of the DDC principle.}
\label{IQprinciple}
\end{center}
\end{figure}

In the second method, the channelized Direct Down Conversion (DDC), the frequency comb is built in real time by summing as many tones as necessary \cite{Swenson}. Then, after propagation of the excitation signal through the MKID array, the resulting signal is processed for each tone by a dedicated DDC (shown in figure \ref{IQprinciple}). The incoming signal, containing all the frequencies of the comb, is at first split in 2 branches and multiplied by the sine and cosine waves of the tone of interest and finally low pass filtered in order to keep only the lower side band, i.e. the baseband, of the signal. The In phase (I) and Quadrature (Q) component of the tone are thus obtained, and with these, the amplitude ($\rm \sqrt{I^2+Q^2}$) and phase ($\rm \arctan (Q/I)$) of each transmitted tones can be continuously monitored.

This method is cumbersome, but provided the Data AcQuisition (DAQ) readout performances meet the throughput, it allows to follow each MKID time-dependent variation up to the low pass filter cut off frequency, which must be at most around one tenth of the mean resonator frequency spacing ($\sim$100\,kHz). The bandwidth, which is here equivalent to the readout rate, can be further narrowed in order to integrate less white noise and thus improving the signal to noise ratio (SNR).
Another benefit is the possibility to modify on-line one, several or all tone frequencies without necessarily stopping the excitation of the untouched tones. 
The maximization of the observation time requires this feature in kilo-pixel camera where the comb frequency must be constantly and promptly adapted to the MKID array resonance shift due to sky condition change (background light, clouds, ...).
This is the approach used and detailed below.

%% file: instrumentation.tex

\section{Experimental setup}
\label{experimentalsetup}
\begin{figure}[th]
\begin{center}
\includegraphics[angle=-90,width=12cm]{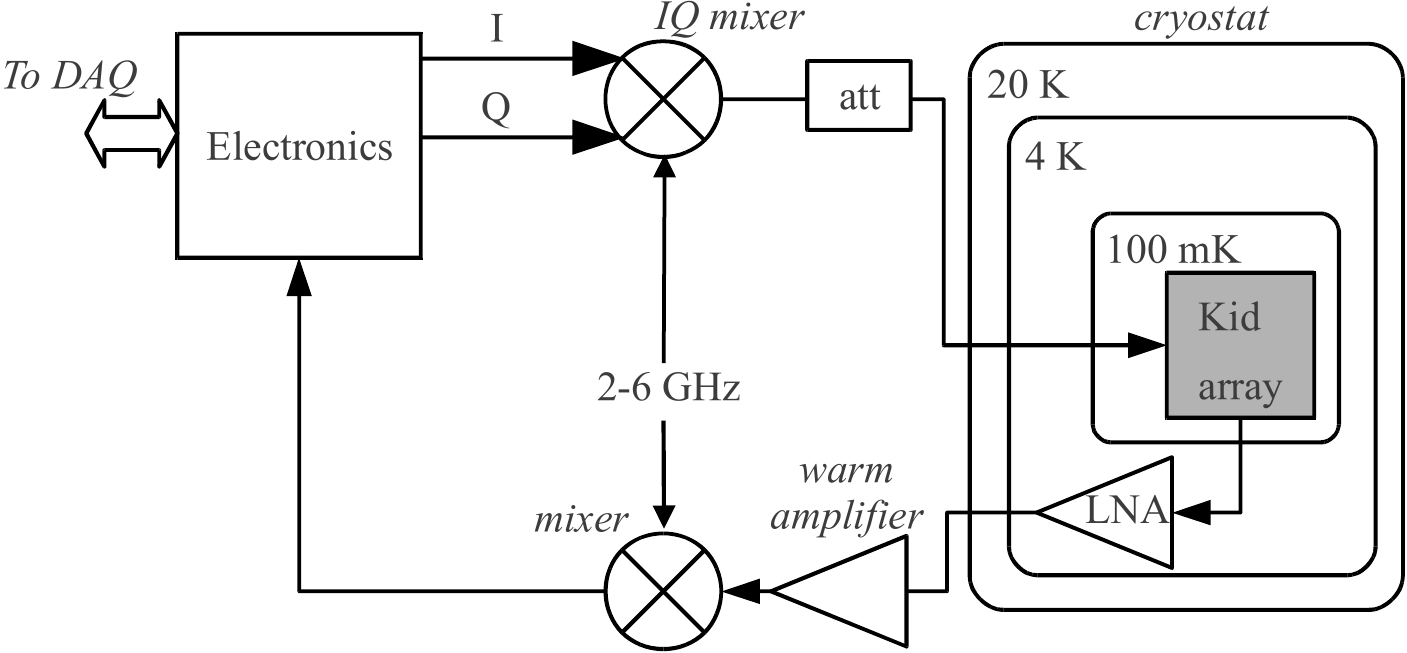}
\caption{Overview of the setup required to monitor a MKID array featuring the electronics generating the two frequency combs (each tone phase shifted by 90\textdegree\  between I and Q), the IQ up-mixer, the programmable attenuator for power adjustment, the cryostat, the low noise boost amplifier, the down mixer and the warm amplifier.}
\label{HardwareSetup}
\end{center}
\end{figure}

Since the MKID resonator frequencies are above several GHz, it is not possible for any electronics using one of the above methods to directly drive the array. Therefore, as shown in figure~\ref{HardwareSetup}, the frequency comb must be generated at baseband and up-mixed to the frequency band of interest by hybrid mixers. The signal is then passed through the array at appropriate power level and finally amplified and down-mixed back to baseband in order to be processed by the electronics. 

The MKID array, operated at 70\,mK, is followed by a low noise cold amplifier (LNA) operated at 4\,K and a warm amplifier. Typically, a combined total gain of 60 to 65\,dB is used to boost the output signal and to adapt it to the ADC input dynamic range of 1\,Vpp. Considering a white noise, the cold amplifier noise temperature of 5\,K reported to the ADC input is equivalent to 1.3\,mV for the considered bandwidth. This yields a best achievable SNR of 57\,dB which is much lower than the 69\,dB SNR of the chosen ADC. Measured performances and noise shape are detailed in section~\ref{ReadoutPerf}.

The programmable attenuator located before the transmission line coupled with the MKIDs adjusts the mean power per tone to an experimentally determined optimum of $\sim$10\,pW.
Note that this value is adapted for typical ground based observations and that space based MKIDs would exhibit much lower saturation powers.
Practical experience shows that driving the resonator with lower power degrades the SNR and using higher power makes the resonator unstable. This noise which is not related to the ``two-level system noise'' described in \cite{GaoTLS} is probably caused by instabilities in the thin superconducting film, and is associated to a strong deformation of the resonance characteristic in the complex plane.

\begin{figure}[th]
\begin{center}
\includegraphics[angle=-90,width=1.0\textwidth]{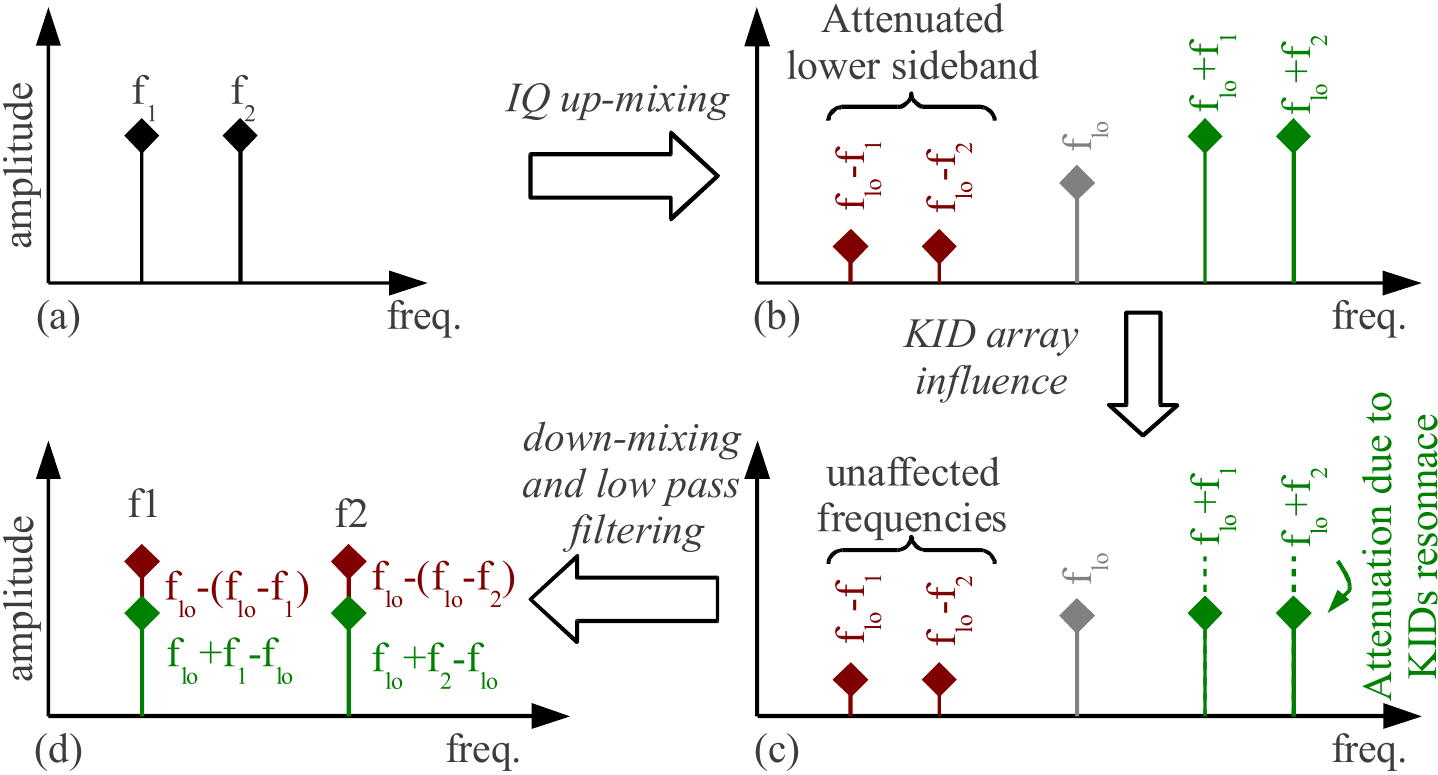}
\caption{Illustration of the up and down-mixing of two tones $\rm f_1$ and $\rm f_2$ by a carrier $\rm f_{lo}$ with ideal mixers (no harmonics generated). 
Part (c) depicts the spectrum affected by the MKID array, only the green part is attenuated by a maximum of 2 to 10\,dB (practical rejection values), i.e it still contains amplitudes 20 to 28\,dB larger than the residuals in red.
As shown on (d), the tones affected by the MKID array and the image frequencies, i.e. the residuals (red), are summed because of the back folding due to the down-mixing.}
\label{MixingIssues}
\end{center}
\end{figure}

It is important to note that the use of an IQ mixer is mandatory, because it presents the benefit to strongly attenuate the lower sideband and the carrier from the output up-mixed spectrum. Typical rejection of -30\,dB of the lower sideband can be obtained. As depicted on figure \ref{MixingIssues}, the spectrum resulting from the down-mixing stage still contains residuals of the lower sideband due to the up-mixing. This part of the spectrum, unaffected by the MKID array, is folded on the tones of interest and reduces their available dynamic range, therefore degrading the SNR.

Another mixer side effect to consider, is their non linearity which creates inevitable intermodulation products. In the up-mixing part, those are not degrading the system performance, since they are either off-resonance or in the worst case tuned to another MKID.  In that case, the considered intermodulation product is only adding a slight extra excitation and therefore is not adding any noise nor crosstalk. In the down-mixing part however, an harmonic of a given MKID resonator frequency mixed to an harmonic of the local oscillator could fall on an other MKID resonance and thus create crosstalk. The probability of such an event is the ratio between the resonance width and their mean separation, i.e. 50\,kHz/2\,MHz=2.5\,\%. Even in that case, this effect is not an issue, since the worst spurious suppression of the Intermediate Frequency harmonics (occurs when M=N in $\rm M \times f_{RF} \pm N \times f_{LO}$) was measured and is better than 45\,dBc  at the operated power.

\section{Hardware development}
\label{hardware}
The complexity of the design is directly driven by the targeted analog bandwidth and the number of tones to manage. The analog bandwidth puts hard constraints on the ADC and on the FPGA. 
On the FPGA side, unless each tone processing is parallelized, the limitation comes from the maximum running frequency of the multipliers used to perform the DDC. The FPGA size directly scales with the number of tones to generate and process.

\begin{figure}[th]
\begin{center}
\includegraphics[angle=-90,width=14cm]{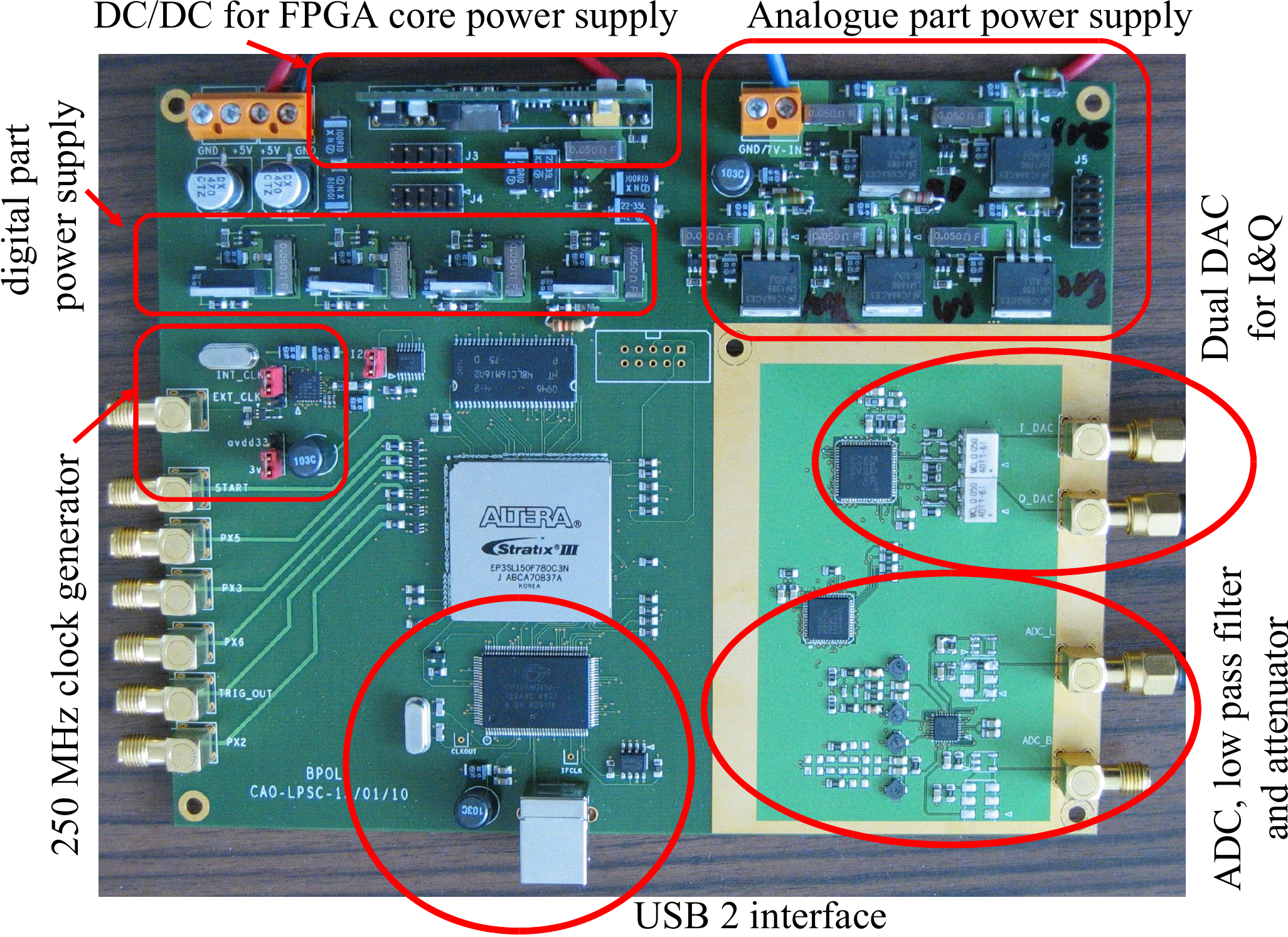}
\caption{Picture of the prototype board.}
\label{BoardPicture}
\end{center}
\end{figure}

The hardware platform is FPGA centered and features a dual DAC for the frequency comb generation, an ADC and a USB2 micro-controller for DAQ and slow control interfacing. A picture of the board can be seen in figure \ref{BoardPicture}. 
In order to demonstrate the concept feasibility and to keep the design simple, it was chosen to limit the sampling and processing frequency to 250\,MHz (analog bandwidth of 125\,MHz). Therefore a 12 bit ADC (AD9230) having an input dynamic range of 1\,Vpp and a SNR of 69\,dB, was used, the excitation being performed by a 14 bit DAC (AD9781). 
The selected FPGA was an ALTERA\textregistered\  stratix III (EP3SL150F780) because it provided a large multiplier count coupled to a large amount of user logic. It can be noted here that the constraints on the FPGA were tightened due to to choice of operating a real-time generation and analysis, as opposed to the FFT method where only fast memory would have been required.\\
The board can be clocked by its local oscillator or by an external reference clock. The latter and the spare trigger inputs are provisions allowing to synchronize several MKID arrays.

\section{Firmware development}
\label{firmware}
\begin{figure}[th]
\begin{center}
\includegraphics[width=14cm]{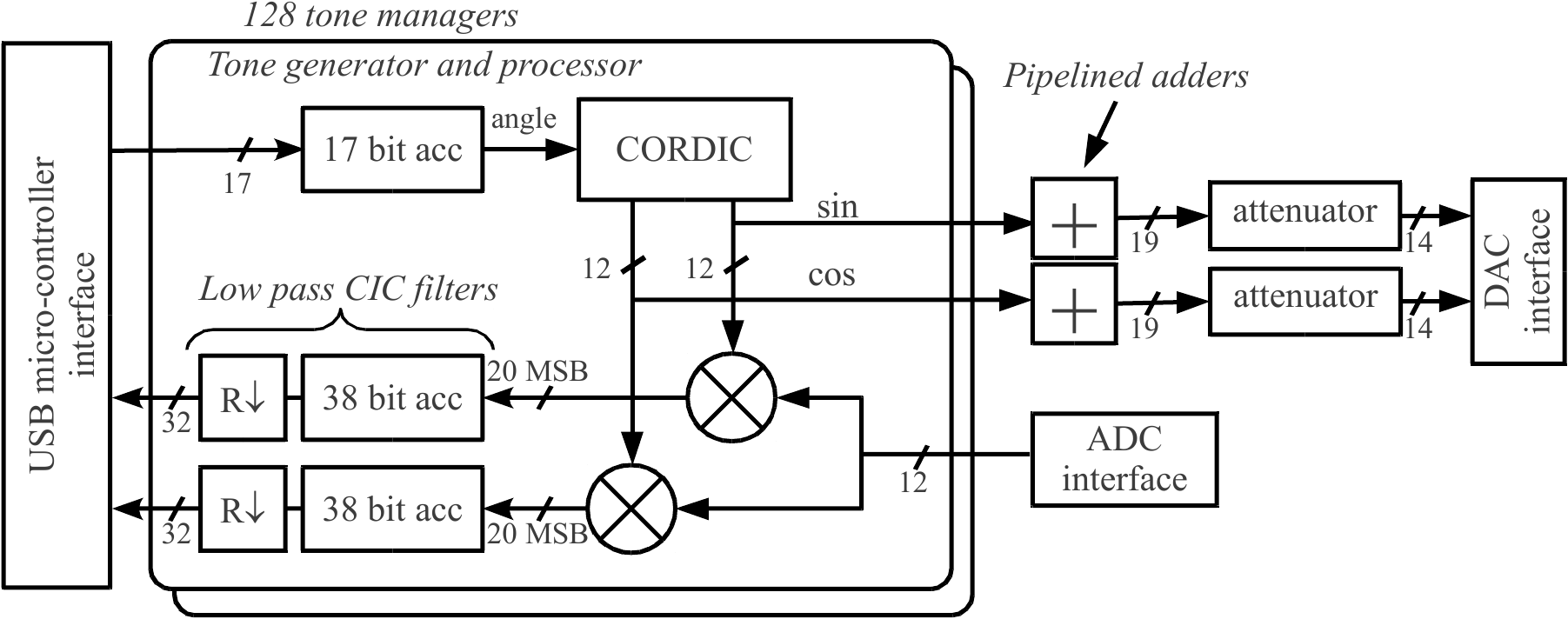}
\caption{Overview of the FPGA firmware}
\label{FPGAContent}
\end{center}
\end{figure}

An overview of the firmware is shown in figure \ref{FPGAContent}. Each tone manager features a COordinate Rotation DIgital Computer (CORDIC) \cite{Volder} block and a DDC that is composed of an I\&Q demodulator followed by a Low Pass Filter (LPF). The LPF, which is primarily used to remove the summed frequencies component from the spectrum, also provides unwanted frequencies rejection (frequencies tuned to other pixels, white noise, \ldots). 

Each CORDIC, implemented in a pipelined fashion and composed only of adders and subtracters, was designed to provide a 12 bit precision on the sine and cosine values calculated. It uses 12 precalculated arc tangent values with 20 bit resolution. The phase accumulator that feeds the CORDIC is used to adjust the frequency with a precision of $\rm 250\ MHz/2^{17} \sim 1.9\ kHz$. The direct and 90\textdegree\  phase shifted 128 tones are summed in parallel by 2 pipelined adders of 19 bit (1 stage per addition) in order to built the in phase and in quadrature version of the comb. To avoid clipping due to the DAC range limitation, the sum result, which is a time periodic signal, must be attenuated (i.e right shifted). Consequently, the dropped bit are monitored and processed by the firmware in order to build an over-range signal whose occurrences in each accumulation frame are counted and returned to the data acquisition. This information is used to compute and apply the optimal shifting. Thus, in the worst case where 128 tones are activated and happen to be regularly summed in phase, the 5 LSB of the result are dropped.
The FPGA largest resources consuming part is the tones generator which uses $\sim$60\% of the available space for 128 generators.

The I\&Q demodulation is performed by multiplying a copy of the ADC output by replicas of the generated sine and cosine values. For practical reasons (FPGA logic resources), the LPF is obtained by averaging $2^{18}$ data and is thus in the order of the kHz of bandwidth. It must be noted, that the accumulator period must be chosen as a multiple of the phase accumulator period in order to avoid beat frequency phenomena. At the end of the accumulation cycle, each tone manager transfers its I\&Q data to the USB interface for readout. That represents a data throughput of $\sim$1\,MB/s for 128 tones. The averaging is then further increased on the readout computer in order to reach a refresh rate of 20\,Hz.

Given the fact that the signal of interest for each pixel is contained in a single tone, the reduction of the bandwidth increases the Signal to Noise Ratio (SNR). Thus, the internal averaging increases the SNR by a factor of $2^{9}$ and the off-line filtering further optimizes it by a factor of $\sim$7. 

\section{Data acquisition and slow control}
\label{dataacquisition}
The computer, the electronics and the power supplies are installed in a 19`` rack. The computer is running a UDP socket server on a linux platform. It provides the interfacing between the USB and Ethernet. The server can receive commands from different computers and does provide data to the clients having subscribed to the acquisition service. Examples of commands are frequency adjustments, gain and attenuation settings.

At the software startup, the FPGA is configured with the current firmware available on the computer and an acquisition thread is started. It reads-out the data and provides on-line analysis features that would have been cumbersome and costly to implement in the hardware, like steeper low pass filtering, first and second order derivative of the I\&Q values (respectively sensitivity and sensitivity variation).
The preprocessed data along with the current settings (frequencies, gain, ...) are sent via indexed datagrams.

%% file: conclusion.tex
\section{Testing}
\label{testing}
The tests performed to assess the performance of the NIKEL multiplexing electronic system are presented in this section.
They were all carried out in laboratory using the cryogenic and sky simulator facilities.

\subsection{Test setup description}

The NIKEL multiplexing electronic system was connected to the currently existing NIKA camera described in the previous sections.
The camera, mounted on a pneumatic table to avoid vibrations, was cooled down to its nominal cryogenic temperature of 70\,mK and electromagnetically
shielded to avoid magnetic field interferences in the MKIDs as well as spill-over radiation from the laboratory floor and walls.
The camera and readout electronics monitoring and control were performed by a dedicated software.

\begin{figure}[th]
\begin{center}
\includegraphics[width=1.0\textwidth]{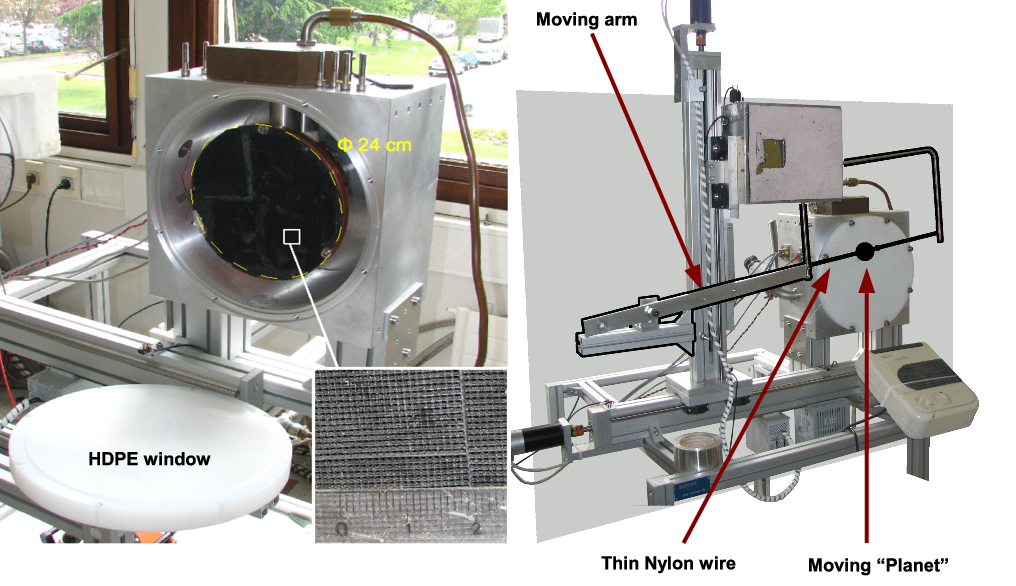}
\caption{Sky simulator used to replicate real observing conditions.}
\label{SkySimulator}
\end{center}
\end{figure}

A fake sky source was observed with the NIKA camera at 150\,GHz in order to test the electronics performance.
For this purpose, the sky simulator (figure \ref{SkySimulator}) which replicates real observing conditions and properly estimate
the amount of stray-light was used.  The basic idea is to cool down a large black disk having the same dimensions as the telescope focal
plane. This cold disk simulates the background temperature in ordinary ground-based observing conditions. The sky simulator
temperature can be continuously adjusted between 50 and 300\,K, allowing an accurate estimation of the detector response
and a direct determination of the noise equivalent temperature (NET). It can also be used to simulate
a point-like source signal to reconstruct the focal plane. This is achieved by placing a room temperature high emissivity ball (``the planet'') in between the camera and the sky simulator cold disk. This ball can be displaced in azimuth and elevation for reproducing
the typical scanning strategy and thus to access the different MKID pixels.

\subsection{Readout performances}
\label{ReadoutPerf}
In order to determine the system noise contributors and thus the readout system limitations, power spectrum distribution (PSD) measurements were performed for three different configurations: direct loopback from the DAC output to the ADC input, loopback in the RF path (cryostat bypassed) and full system with the MKID array and the cold amplifier. In the later case, the excitation signal contained on and off resonances frequencies.
\begin{figure}[th]
\begin{center}
\includegraphics[angle=-90,width=0.8\textwidth]{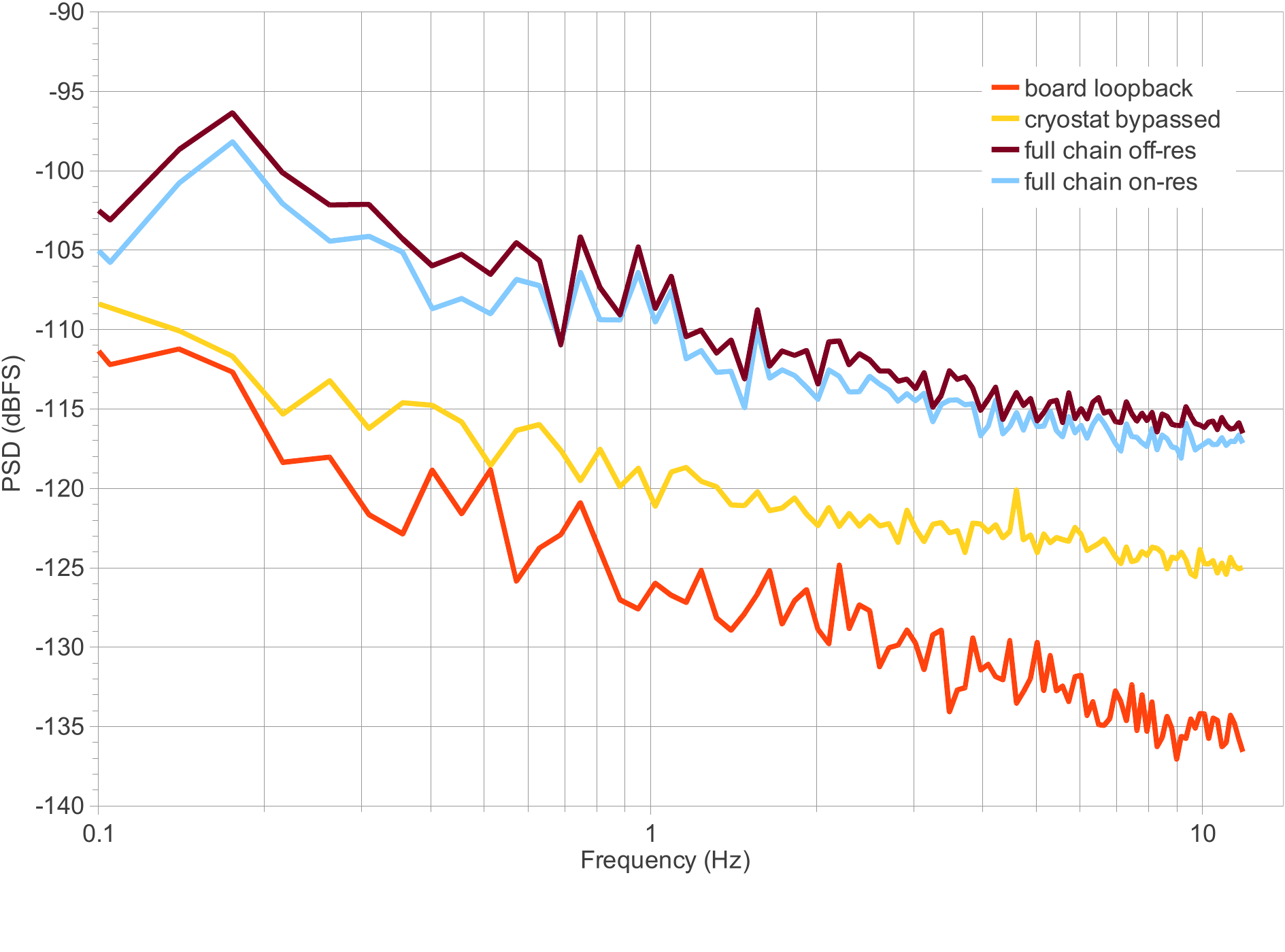}
\caption{Power spectrum distribution plot showing the different contributors to the system noise floor.}
\label{NoiseContributor}
\end{center}
\end{figure}
Figure~\ref{NoiseContributor} shows that the SNR is degraded by a 1/f noise due to the readout electronic and that it is experienced as well by on and off resonance frequency. Therefore this noise contribution, which is highly correlated, can be mostly removed by off-line processing. Moreover, the plot indicates that the full system noise floor is reached at -116\,dB, while it is reached at -135\,dB for the readout electronics alone. This demonstrate that the performance limiting factor is mostly due to the cold amplifier noise temperature.

\begin{figure}[th]
\begin{center}
\includegraphics[angle=-90,width=0.85\textwidth]{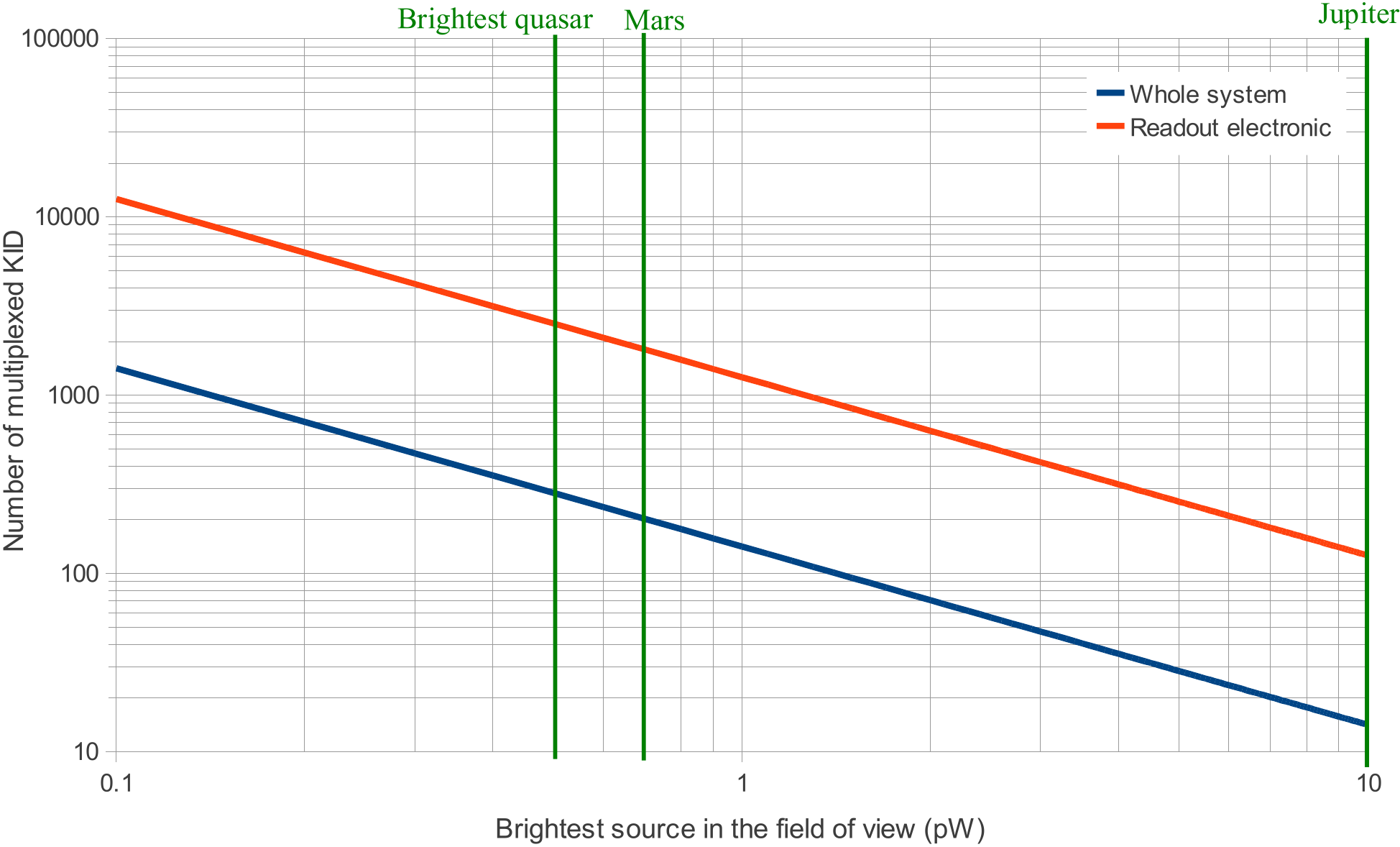}
\caption{Readout system multiplexing limit expressed as a function of the brightest source in the field of view (FOV) with a minimum resolution fixed at the photon noise limit. Common sources power are provided on the plot. With the brightest quasar in the FOV, a maximum of 200 MKIDs can be multiplexed in a single line.}
\label{MultiplexingLimit1}
\end{center}
\end{figure}

The multiplexing limitation of the system can be determined by the following relation:
\begin{equation}
\rm  D > 20 \times \log (\frac{P}{S} \times N)
\end{equation}
Where D is the necessary dynamic range in order to instrument N MKIDs with an individual resolution S over a range P. Considering the available dynamic range of 116\,dB (see figure~\ref{NoiseContributor}), the plot in figure~\ref{MultiplexingLimit1} shows the multiplexing limit achievable as a function of the brightest source in the field of view, using a fixed resolution S. The photon noise limit of about $\rm 50\ aW/\sqrt{Hz}$, over a representative bandwidth of 20\,Hz is used as the target resolution.

\subsection{Optical results}
\begin{figure}[th]
\begin{center}
\includegraphics[angle=-90,width=0.8\textwidth]{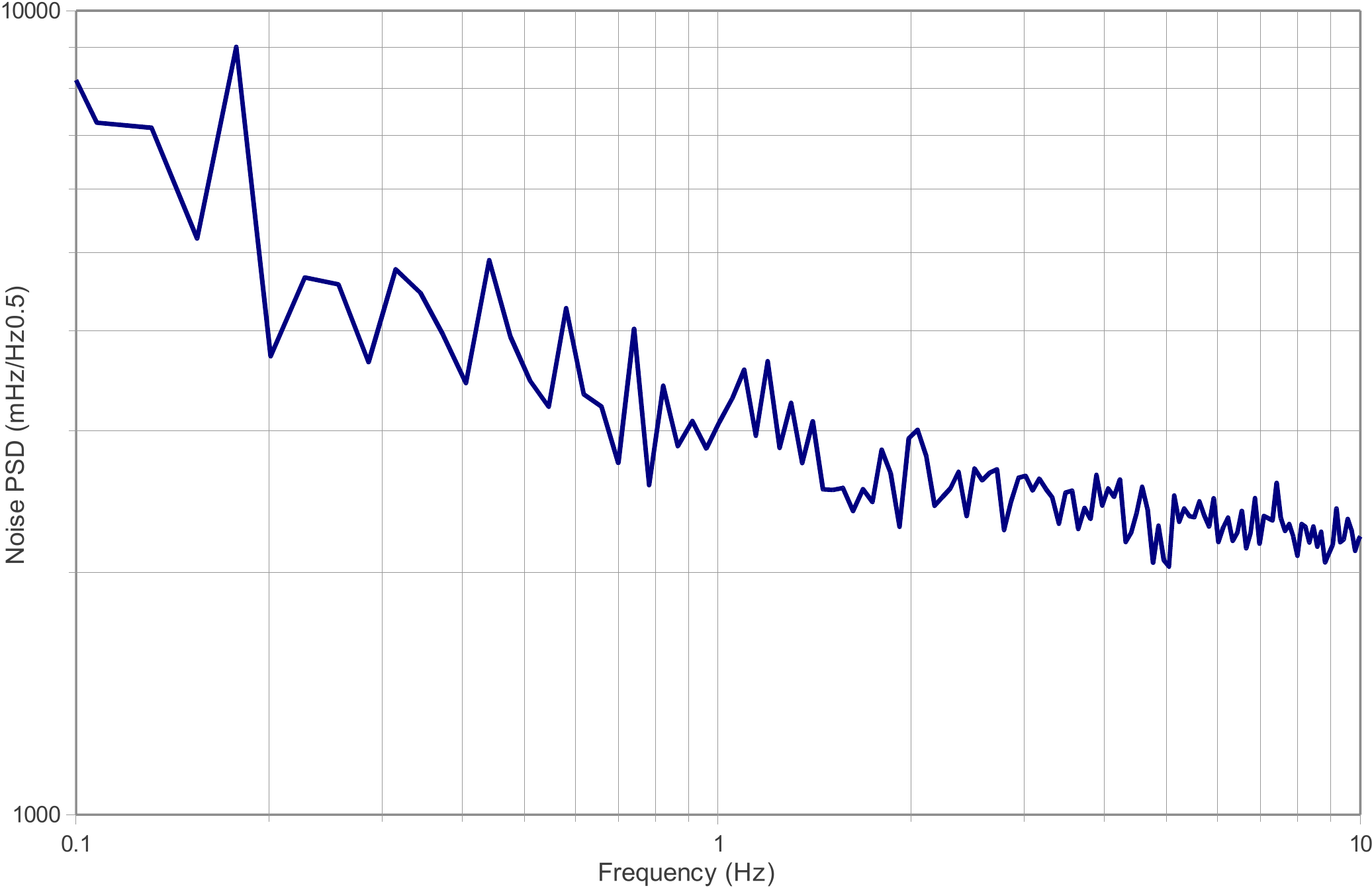}
\caption{Average frequency noise power spectrum in units of $\rm \frac{mHz}{\sqrt{Hz}}$.}
\label{noiseroche}
\end{center}
\end{figure}

Figure~\ref{noiseroche} shows the averaged noise power spectrum in units of $\rm \frac{mHz}{\sqrt{Hz}}$ for 64 MKID detectors, spaced by $\sim$2\,MHz, which were simultaneously monitored. 
This roughly corresponds to the maximum possible number of detectors as given by the electronic equivalent bandwidth, i.e. 125\,MHz.
At the representative frequency of 1\,Hz, the noise is about 3$\rm \frac{Hz}{\sqrt{Hz}}$, so displacements of the resonance frequency of the order of a few Hz are detectable in one second integration time.
As explained in a previous section, the observed noise level, which is comparable to a single detector readout system, is actually dominated by the cold amplifier and not by the NIKEL electronics. 
From this, it can be concluded that the NIKEL electronics can be used for multiplexing with no performance loss.

\begin{figure}[th]
\begin{center}
\includegraphics[width=14cm]{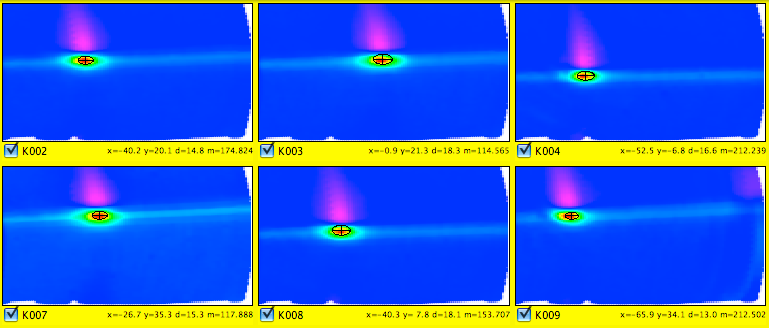}
\caption{Focal plane reconstruction using the sky simulator for 6 pixels. Each individual pixel map exhibits a well
defined beam, fitted with a 2D gaussian (FWHM projected circle shown).}
\label{lobes1}
\end{center}
\end{figure}

Figure~\ref{lobes1} shows images of the sky simulated source (the so-called planet) for six detectors.
The negative vertical pattern in the figure is an effect of the off-line processing high-pass filtering and the horizontal line is the image of the thin nylon wire holding the ball.
The plots show that the simulated source is well centered at the expected pixel position. Also no spurious features are
found on the other pixels which are read simultaneously by the electronic system, i.e. no crosstalk effect.

\begin{figure}[th]
\begin{center}
\includegraphics[angle=-90,width=0.8\textwidth]{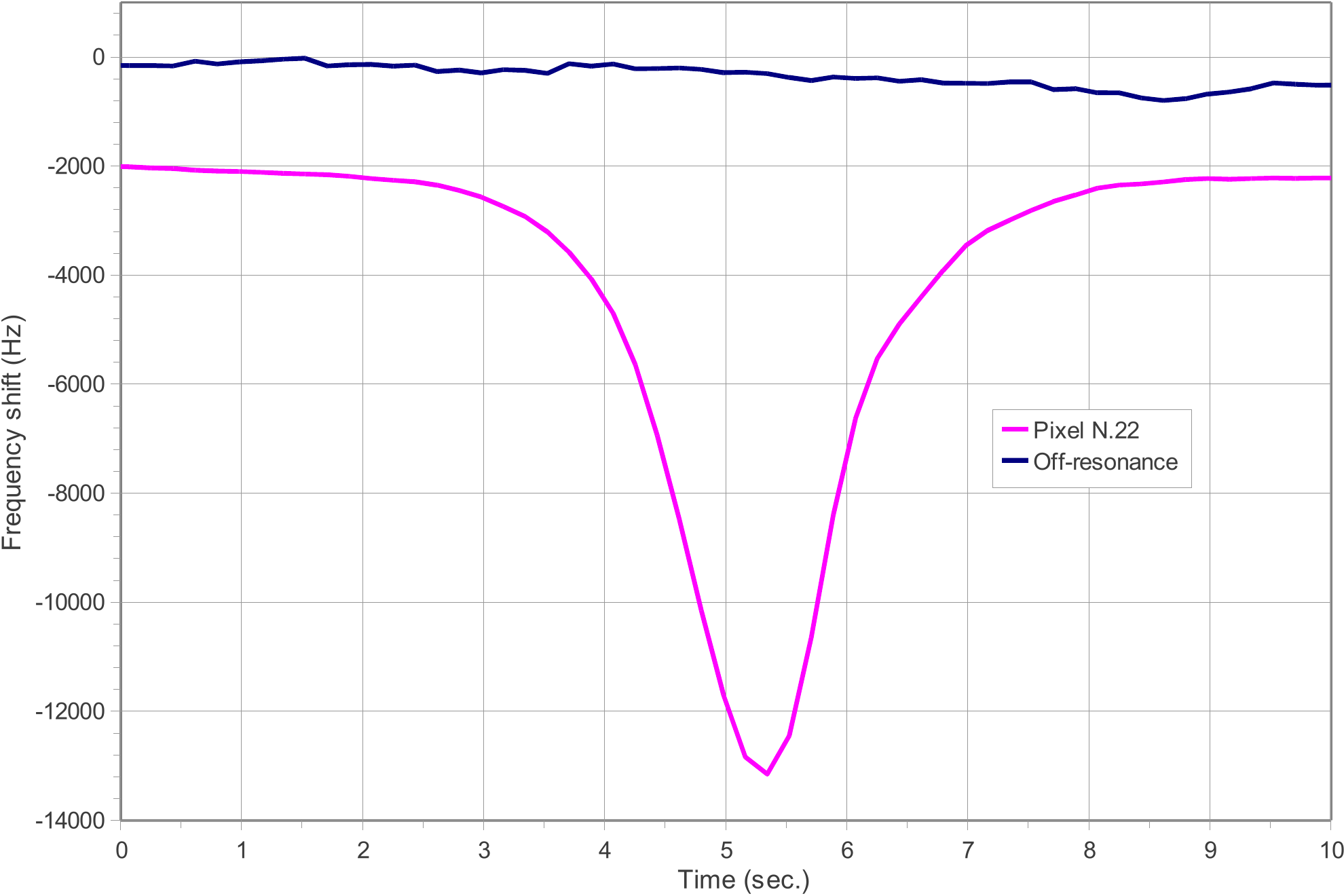}
\caption{Time-domain trace of one sky simulator planet passage. An optically active pixel is plotted along with one off-resonance blind tone. Nothing is detected on the blind, despite the large signal on the adjacent electronic channel.}
\label{crosstalk}
\end{center}
\end{figure}

To further stress the fact that the cross-talk potentially introduced by the readout electronics is negligible, we plot in figure~\ref{crosstalk} a passage of the sky simulator fake planet (equivalent to mars) over a pixel, compared with the simultaneous readout of one of the adjacent tones, placed off-resonance. Despite the large shift in frequency of the on-resonance tone (about 12\,kHz), nothing is detected on the blind electronics channel. We place in this way an upper limit of about one percent on the total cross-talk, totally accepted for any astronomical observation in the mm-wave band.

\section{Conclusions and perspectives}
\label{conclusion}
As the results show, real-time frequency multiplexing is a promising method of instrumentation for the MKID arrays. Future developments should focus on further extending the useful bandwidth and the number of tones processed in order to permit the equipment of kilo-pixels scale camera. Another path of progression is to implement individual tone amplitude adjustment, in order to select the optimum applied power per resonator.